\documentclass{aa}  

\usepackage{epsfig,graphicx,mathptmx}
\usepackage{amsmath}
\usepackage[T1]{fontenc}
\usepackage{amssymb}
\usepackage[breaklinks,colorlinks,citecolor=blue]{hyperref}
\usepackage{enumerate}
\usepackage{txfonts}
\def\stacksymbols #1#2#3#4{\def\theguybelow{#2}
        \def\verticalposition{\lower#3pt}
        \def\spacingwithinsymbol{\baselineskip0pt\lineskip#4pt}
        \mathrel{\mathpalette\intermediary#1}}
\def\intermediary #1#2{\verticalposition\vbox{\spacingwithinsymbol
        \everycr={}\tabskip0pt
        \halign{$\mathsurround0pt#1\hfil##\hfil$\crcr#2\crcr
                \theguybelow\crcr}}}

\def\kms{{\rm km\:s^{-1}}}

\newcommand{\sqr}{$^2$}
\newcommand{\cub}{$^3$}
\newcommand{\lcdm}{$\Lambda$CDM}
\newcommand{\omegam}{$\Omega_{\rm m}$}

\newcommand{\omegal}{$\Omega_\Lambda$}

\newcommand{\tsl}{$T_{\rm sl}$}
\newcommand{\vew}{$v_{\rm ew}$}
\newcommand{\wew}{$w_{\rm ew}$}
\newcommand{\hzero}{$H_0$}
\newcommand{\msun}{$M_{\sun}$}
\newcommand{\zsun}{$Z_{\sun}$}
\newcommand{\ptd}{$P_{\rm 2D}$}
\newcommand{\mn}{$\mathcal{M}$}
\newcommand{\mnall}{$\mathcal{M}=(0.25,0.5,0.75)$}

\newcommand{\apec}{\textsc{apec}}
\newcommand{\bapec}{\textsc{bapec}}
\newcommand{\xspec}{\textsc{xspec}}
\newcommand{\sixte}{\textsc{sixte}}

\DeclareMathAlphabet{\mathcal}{OMS}{cmsy}{m}{n}

\def\aap{A\&A}

\def\apjl{ApJ}

\def\apjs{ApJS}

\begin{document} 

\title{Measuring turbulence and gas motions in galaxy clusters via synthetic {\it Athena} X-IFU observations}
\titlerunning{Measuring gas motions in clusters with {\it Athena} X-IFU}
\authorrunning{M. Roncarelli et al.}

\author{
M.~Roncarelli\inst{1,2} \and
M.~Gaspari\inst{3}\fnmsep\thanks{{\it Einstein} and {\it Spitzer} Fellow.} \and
S.~Ettori\inst{2,4} \and
V.~Biffi\inst{5,6} \and
F.~Brighenti\inst{1} \and
E.~Bulbul\inst{7} \and
N.~Clerc\inst{8} \and
E.~Cucchetti\inst{8} \and
E.~Pointecouteau\inst{8} \and
E.~Rasia\inst{6}
}

\institute{
Dipartimento di Fisica e Astronomia, Universit\`a di Bologna, via Gobetti 93 I-40127 Bologna, Italy \\
\email{mauro.roncarelli@unibo.it}
\and
Istituto Nazionale di Astrofisica (INAF) -- Osservatorio di Astrofisica e Scienza dello Spazio (OAS), 
via Gobetti 93/3, I-40127 Bologna, Italy
\and
Department of Astrophysical Sciences, Princeton University, 4 Ivy Lane, Princeton, NJ 08544-1001, USA
\and
Istituto Nazionale di Fisica Nucleare (INFN) -- Sezione di Bologna, viale Berti 
Pichat 6/2, I-40127 Bologna, Italy
\and
Dipartimento di Fisica dell'Universit\`a di Trieste, Sezione di Astronomia, via Tiepolo 11, 
I-34131 Trieste, Italy
\and
Istituto Nazionale di Astrofisica (INAF) -- Osservatorio Astronomico di Trieste, via Tiepolo 11, 
I-34131, Trieste, Italy
\and
Harvard-Smithsonian Center for Astrophysics, 60 Garden Street, Cambridge, MA, 02138, USA
\and
IRAP, Universit\'e de Toulouse, CNRS, UPS, CNES, 31042 Toulouse, France \\
}

   \date{Received 5 May 2018 / Accepted 3 July 2018}

 
  \abstract
   {The X-ray Integral Field Unit (X-IFU) that will be on board  the 
   {\it Athena} telescope will provide an unprecedented view of the 
   intracluster medium (ICM) kinematics through the observation of gas 
   velocity, $v$, and velocity dispersion, $w$, via centroid-shift and 
   broadening of emission lines, respectively.}
   {The improvement of data quality and quantity requires an assessment 
   of the systematics associated with this new data analysis, namely 
   biases, statistical and systematic errors, and possible correlations 
   between the different measured quantities.}
   {We have developed an end-to-end X-IFU simulator that mimics a full 
   X-ray spectral fitting analysis on a set of mock event lists, obtained 
   using \sixte. We have applied it to three 
   hydrodynamical simulations of a Coma-like cluster 
   that include the injection of turbulence. This allowed us to assess the 
   ability of X-IFU to map five physical quantities in the cluster core: 
   emission measure, temperature, metal abundance, velocity, and velocity 
   dispersion. Finally, starting from our measurements maps, we computed the 
   2D structure function (SF) of emission measure fluctuations, $v$ and $w$, 
   and compared them with those derived directly from the simulations. }
   {All quantities match with the input projected values without bias; the  systematic errors 
   were below 5\%, except for velocity dispersion whose error reaches about 15\%. 
   Moreover, all measurements prove to be statistically independent, indicating 
   the robustness of the fitting method. Most importantly, we recover the 
   slope of the SFs in the inertial regime with excellent accuracy, but we 
   observe a systematic excess in the normalization of both SF$_v$ and SF$_w$ 
   ascribed to the simplistic assumption of uniform and (bi-)Gaussian 
   measurement errors.}
   {Our work highlights the excellent capabilities of {\it Athena} X-IFU in
   probing  the thermodynamic and kinematic properties of the 
   ICM. This will allow us to access  the physics of its 
   turbulent motions with unprecedented precision.}


   \maketitle
%


\section{Introduction} \label{s:intro}

The Advanced Telescope for High ENergy Astrophysics\footnote{\href{http://www.the-athena-x-ray-observatory.eu}{http://www.the-athena-x-ray-observatory.eu}} ({\it Athena}) 
is the next-generation, groundbreaking X-ray observatory mission selected by ESA to study the Hot and Energetic Universe; its   launch is expected in 2028 (\citealt{Nandra:2013}).  With its two instruments, the Wide Field Imager and the X-ray Integral Field Unit \citep[X-IFU;][]{barret16}, {\it Athena} will provide a crucial window on the X-ray sky, strongly complementing the other next-generation telescopes covering the lower energy bands (e.g.,~ALMA, ELT, Euclid, JWST, SKA).
In particular, the X-IFU will consist of a cryogenic X-ray spectrometer with a large array of transition-edge sensors, with the goal of achieving 2.5 eV spectral resolution, with 5'' pixels, over a field of view (FOV) of about 5 arcmin. The X-IFU is expected to deliver superb spatially resolved high-resolution spectra, opening the era of imaging spectroscopy in the X-ray band.

One of the primary goals of {\it Athena} X-IFU is probing the (astro)physics of the diffuse plasma which fills the potential wells of galaxy clusters (\citealt{Ettori:2013,Croston:2013,pointecouteau13}),  also known as the intracluster medium (ICM). The ICM, which formed during the cosmological gravitational collapse in the growing dark matter halo (e.g.,~\citealt{Planelles:2017}),  accounts for most of the baryonic mass of a massive galaxy cluster ($\sim 12-15$ \%\ of the total mass of $\sim 10^{15}$\,\msun), with typical electron temperatures and densities in the core of $T_{\rm e}\sim 5-10$ \,keV and $n_{\rm e}\sim 10^{-2}-10^{-3}$\,cm$^{-3}$, respectively. While the thermodynamics of the ICM,   particularly in the core region ($<0.1\,R_{\rm vir}$),  has been fairly well constrained by XMM-{\it Newton} and {\it Chandra}, its kinematics remains very poorly understood.

On the theoretical side, cosmological simulations show that the ICM is continuously stirred by the accretion of substructures or merger events at 100s kpc scale (e.g.,~\citealt{dolag05,vazza11b,Lau:2009,Lau:2017}) and by the self-regulated active galactic nucleus (AGN) feedback in the inner core (e.g.,~\citealt{GS:2017} for a brief review of the self-regulated cycle). In both cases, the injected turbulence is subsonic in the ICM with a few 100\,$\kms$ magnitude. Until recently the observational constraints on ICM gas motions have been indirectly derived from measurements of X-ray surface brightness fluctuations (e.g.,~\citealt{Schuecker:2004,churazov12,Gaspari:2013_coma,zhuravleva15,walker15,Hofmann:2016}), resonant scattering (e.g.,~\citealt{churazov04,Ogorzalek:2017,hitomi18}), and Sunyaev--Zeldovich fluctuations \citep{Khatri:2016}. At the same time, XMM-{\it Newton} RGS grating observations have shown the potential for high-resolution spectroscopy to directly infer gas motions from the spectral line broadening, setting significant upper limits on the ICM turbulence in several cool-core clusters \citep[e.g.,][]{sanders10,bulbul12,Sanders:2013,Pinto:2015}.
In its brief life, only the Soft X-ray Spectrometer Calorimeter on board the {\it Hitomi} satellite has given us the first direct detection of the  line-of-sight ICM velocity dispersion in the inner 100 kpc of the Perseus cluster, showing turbulent velocities on the order of $150\,\kms$ (\citealt{Hitomi:2016,hitomi18}). A similar level of chaotic motions has been confirmed via the well-resolved velocity dispersion of the ensemble H$\alpha$+[NII] warm gas condensing out of the turbulent ICM (\citealt{Gaspari:2018}). 

Given the vastly improved capabilities compared with its predecessor XMM-{\it Newton} RGS (see, e.g.,~\citealt{Sanders:2013}), X-IFU will be able to directly test turbulent and bulk motions via two observables: the broadening and the shift of the centroids of the emission lines in the X-ray band (such as the Fe, L, and K complexes; e.g.,~\citealt{Inogamov:2003}). This will allow a direct measurement of motions that are known to drastically affect the physics of the ICM inducing cosmic ray re-acceleration (e.g.,~\citealt{brunetti01,petrosian01,Eckert:2017_rad,bonafede18}), thermal instability (e.g.,~\citealt{Gaspari:2015,voit18}), and magnetic field amplification (e.g.,~\citealt{cho09,santos-lima14,Beresnyak:2016}). Moreover, the understanding of ICM kinematics, combined with the more traditional techniques of density and temperature mapping, has the potential to improve the robustness of cluster mass measurements. While current X-ray analyses have to rely on the  hydrostatic equilibrium hypothesis, it may lead to underestimates on the order of 10--20\% (see, e.g.,~\citealt{piffaretti08,Lau:2009,rasia12,shi16,biffi16}), new constraints on the kinematics of cluster plasma will allow us to test them directly. Finally, these new observables will improve our knowledge of cluster formation and structure assembly up to the virial regions where the ICM becomes more inhomogeneous and connects with the large-scale structure of the Universe \citep[see, e.g.,][]{roncarelli06b,roncarelli13,vazza11,eckert15,Eckert:2018}.

The  introduction of imaging spectroscopy, with its  increased amount of information, will require the definition of new methods of X-ray data analysis. A crucial point in order to fully exploit this large set of high-resolution spectra is the assessment of biases and systematics that may arise from instrumental effects and limits of the X-ray fitting methods. In this work, we anticipate the key impact of {\it Athena} X-IFU by running for the first time a series of end-to-end advanced synthetic observations, starting from hydrodynamical simulations, carefully including all observational effects (i.e.,~projection, noise, instrumental background) down to the measurement of all physical quantities via spectral fitting. In this exploratory investigation we emphasize the intrinsic physical quantity definitions, and we investigate possible systematic biases and uncertainties on their retrieval as well as statistical correlations of their measurements. All these properties must be first precisely assessed in order to provide robust physics estimates on the nature of turbulence and bulk motions that remain the ultimate goals of X-IFU observations. Overall, this study will be invaluable to fully leverage the X-ray capabilities of the next-generation {\it Athena} observatory, at the same time helping the community to find the more effective path to advance the X-ray ICM field in the next decade.

The paper is structured as follows. Section~\ref{s:models} reviews the high-resolution hydrodynamical 3D simulations used in this study and how we model the {\it Athena} X-IFU synthetic observations. Section~\ref{s:results} presents the results of the synthetic data analysis, while carefully dissecting all the biases and uncertainties. In Section~\ref{s:sf}, we  make extensive use of structure functions (mathematically tied to power spectra) to assess the main ICM kinematics and thermodynamics at varying scales. Finally, we summarize and present concluding remarks in Section~\ref{s:concl}. Throughout  the paper we assume a flat \lcdm\ cosmological model, with \omegam\,=\,0.3, \omegal\,=\,0.7, and \hzero\,$=70$ km\,s$^{-1}$\,Mpc$^{-1}$. When quoting metal abundances, we refer to the solar 
value (\zsun) as measured by \cite{anders89}. To avoid confusion with the error quotations, we use $v$ to indicate velocities and $w$ (instead of  $\sigma_v$, for example) for velocity dispersions, and refer to their errors as $\sigma_v$ and $\sigma_w$, respectively. Quoted errors indicate $\pm1\sigma$ uncertainties.


\section{Models and method} \label{s:models}

\subsection{High-resolution hydrodynamical simulations}

The starting point of our work is the output of the high-resolution hydrodynamical simulations 
of galaxy clusters presented in \citet{Gaspari:2013_coma} and \citet{Gaspari:2014_coma2} (hereafter G13 and G14). Here 
we briefly summarize their main features and ingredients. We refer the interested reader to G13 
for the details and in-depth discussions of the related numerics and physics. 

The G13 and G14 simulations have been structured to be controlled experiments of key plasma 
physics in an archetypal massive and hot galaxy cluster similar to  the  Coma cluster, with virial 
mass $M_{\rm vir}\approx10^{15}\,M_\odot$ and $R_{\rm vir}\approx2.9$ Mpc 
($R_{\rm 500}\approx1.4$ Mpc). The initial density and temperature profiles of the hot 
plasma are set to match deep XMM-{\it Newton} observations of Coma, which shows characteristic central 
density $n_{\rm e, 0} \simeq 4\times10^{-3}$ (with beta profile index $\beta=0.75$) and 
flat core temperature $T_0=8.5$\,keV (with declining large-scale profile 
$\propto r^{-1}$). Given the large core temperature, Coma is the ideal laboratory for testing 
the effect of plasma transport mechanisms as thermal conduction.
The 3D box has a side length $\simeq 1.4$\,Mpc. We adopted a uniform, fixed grid of $512^3$ to 
accurately study perturbations and turbulence throughout the entire volume, reaching a spatial 
resolution $l_{\rm pix} \simeq 2.7$\,kpc.

By using a modified version of the Eulerian grid code \texttt{FLASH4} (with third-order unsplit 
piecewise parabolic method), we solve the 3D equations of hydrodynamics for a realistic 
two-temperature electron-ion plasma, testing different levels of turbulence injection and electron 
thermal conduction (G13 Eqs.~3--7). The plasma is fully ionized with atomic weight for 
electrons and ions of $\mu_{\rm i}\simeq1.32$ and $\mu_{\rm e}\simeq1.16$, respectively, 
providing a total gas molecular weight $\mu\simeq0.62$. The ions and electrons equilibrate via Coulomb 
collisions on a timescale $\gtrsim 50$\,Myr in the diffuse and hot ICM (G13 Sec.~2.5).

Turbulence is injected in the cluster via a spectral forcing scheme based on an 
Ornstein--Uhlenbeck random process (G13 Sec.~2.3), which reproduces experimental 
turbulence structure functions (\citealt{Fisher:2008}).
The stirring acceleration is first computed in Fourier space and then directly converted 
to physical space. Modeling the action of mergers and cosmic inflows (e.g.,~\citealt{Lau:2009, Miniati:2014}), 
the injection scale is set to have a peak at 
$L\approx600$\,kpc, thus exploiting the full dynamical range of the box. By setting the 
energy per mode, we can control the injected turbulence velocity dispersion or 3D Mach 
number, \mn$=w/c_{\rm s}$, which is observed to be subsonic in the ICM (e.g.,~\citealt{Pinto:2015,Khatri:2016}). 
We  considered three different values of the Mach number: 
\mn\,=\,0.25, 0.50, 0.75 (the Coma sound speed is $c_{\rm s}\approx1500\,\kms$). 
We evolve the system for at least two eddy turnover times ($\sim$\,2 Gyr for the slowest models), 
$t_{\rm turb}\sim L/w$, to achieve a statistical steady state of the turbulent cascade. The 
turbulent diffusivity at injection can be defined as $D_{\rm turb}\approx w\,L$.

In this paper, we only analyze  the hydro runs with highly suppressed conduction ($f=0$). 
Recent constraints, for example  the survival of ram-pressure stripped group tails 
(\citealt{DeGrandi:2016,Eckert:2017_cond}) and the presence of significant relative 
density and surface brightness fluctuations in the bulk of the ICM (e.g.,~G13; \citealt{Hofmann:2016, 
Eckert:2017_rad}), indicate a high level of suppression ($f\lesssim 10^{-3}$) compared 
with the classic \citet{Spitzer:1962} value. Highly tangled magnetic fields and plasma 
micro-instabilities (e.g., firehose and mirror) conspire to reduce the electron (and 
ion) mean free path well below the collisional Coulomb scale.

\subsection{Modeling X-IFU observations}

In this section we describe our synthetic observation method, which ultimately generates
a set of realistic mock X-IFU event lists. The first step consists in creating for every 
hydrodynamical simulation an $(x,y,E)$ spectroscopic data cube  that contains the ideal 
imaging and spectra of photons that impact the telescope (i.e.,~without instrumental effects) 
as a function of sky position $(x,y)$ and energy, $E$ \citep[see a similar 
procedure described in][]{roncarelli12}. In a second phase we use this 
information as an input for the SImulation of X-ray TElescopes \citep[\sixte,][]{wilms14} 
code, the official simulator of the \emph{Athena} instruments. From that we finally obtain 
a set of mock X-IFU event lists.

Starting from the outputs of the hydrodynamical code, we model the expected X-ray emission 
of each simulated $l_{\rm pix}^3$ cell. We place our simulated clusters at a comoving distance 
corresponding to $z_0=0.1$ and consider the telescope to be oriented along one of the simulation 
axes to facilitate the line-of-sight integration. With this configuration the 1.4 Mpc cube width 
corresponds to 12.6 arcmin, so that the width of the X-IFU FOV ($\sim$5 arcmin) corresponds to 
 $\sim$40 \%\ of the cube side. On the other hand, each volume element is 1.5 arcsec wide: 
this means that considering the \sixte\ detector configuration, an X-IFU pixel (4.38 arcsec wide) 
encloses on average $\sim$8.5 different simulation cells, i.e.,~more than 4000 volume 
elements when considering the integration along the line of sight.

For each of the 512\cub\ cubic cells we consider its density $\rho_i$, its (electron) 
temperature $T_i$, and its velocity $v_i$ along the line of sight  provided by 
the simulation, and use them to compute the expected emission spectra assuming an \apec\ 
model \citep[version 2.0.2,][]{smith01}. In detail, we use the \xspec\ software (version 
12.9.0i) and fix the effective redshift to
\begin{equation}
z_i = (1+z_0) \sqrt{\frac{1+\frac{v_i}{c}}{1-\frac{v_i}{c}}} -1 \, ,
\label{e:zs}
\end{equation}
being $c$ the speed of light in vacuum. The normalization\footnote{Here we use the 
\xspec\ convention for the normalization, which includes the $10^{-14}$ factor and is 
expressed in units of cm$^{-5}$.} is then fixed to
\begin{equation}
\mathcal{N}_i = \frac{10^{-14} \, n_{e,i} \ n_{{\rm H},i} \ l_{\rm pix}^3}{4\pi \, d_c^2(z_i)} \, 
\label{e:norm}
,\end{equation}
being $d_c(z_i)$ the comoving distance, and where $n_{e,i}$ 
and $n_{{\rm H},i}$ are the electron and hydrogen number density derived from $\rho_i$ 
assuming a primordial He mass abundance of $Y=0.24$. Since our hydrodynamical simulations 
do not provide a self-consistent metal enrichment scheme, our spectra are computed 
assuming a uniform metal abundance of $Z$=0.3 \zsun. We also fix the Galactic hydrogen 
column density to $N_{\rm H}=5\times 10^{20}$ cm$^{-2}$, which corresponds to an average 
absorption value. After computing the spectra for every cell, we obtain the 
$(x,y,E)$ data cube by integrating the spectra along the line of sight. This is done for 
each of the three simulation runs. Regarding the properties of emission-lines, which is  the most 
important aspect of our work, we note that 
while thermal broadening is modeled by the \apec\ model itself and is derived from the cell 
temperature $T_i$, velocity or turbulence broadening is not a direct input of our 
simulations; instead, it emerges in the projection phase as a result of the superimposition of 
spectra with different peculiar velocities $v_i$  provided by the hydrodynamical 
simulations. Our approach, therefore, mimics the real physical process without introducing any 
further assumptions. We fix the energy binning of our $(x,y,E)$ data cubes to 1 eV 
uniformly from 2 to 8 keV, enough to sample the expected 2.5 eV X-IFU energy resolution.

This set of data is then used as  input for the \sixte\ software. Following the specifications of the \sixte\ manual\footnote{http://www.sternwarte.uni-erlangen.de/research/sixte/} 
\citep[see also][]{schmid13}, each data cube is assumed to be placed in a given 
position in the sky, defined by the coordinate of its center. This allows us to define a 
set of sources that cover the  full FOV of X-IFU,  each of which has a proper spectrum. These 
sources are then processed by the \sixte\ software, assuming 
the most recent X-IFU characteristics, to provide a mock observed  event list. Specifically, starting from the input spectra \sixte\ performs a Monte Carlo generation of photons,  
assumed to impact the telescope with given position, energy, and time. Each mock 
photon is then processed to derive whether it is detected, and if so the position and 
energy measured by the detector are saved. This process considers the PSF of the instrument, its 
response function, the vignetting due to the telescope optics, and other detector effects 
such as its geometry, filling-factor, cross-talk, and pile-up. A more detailed explanation 
of the implementation of these effects can be found in the \sixte\ manual and paper \citep{wilms14}. 

In order to reach the same statistical significance, we vary the exposure time 
$t_{\rm exp}$ for each of the three simulated observations to obtain a total number of 
$N_{\rm ph}=18 \times 10^6$ photons in the 2--8 keV energy range. Given 
the different thermodynamical properties of the simulations (see~G13), this translates into a factor of 
$\sim2.5$ difference in $t_{\rm exp}$ between the 
\mn\,=\,0.75 and the \mn\,=\,0.25 simulations. We verified that in order to reach this number of 
photon counts with a $T\sim 8$ keV cluster at $z=0.1$ we would need $t_{\rm exp} \simeq 
2$ Ms, thus significantly higher with respect to the predictions of realistic 
X-IFU observations. On the other hand, this allows us to  substantially reduce the resulting statistical 
errors and to provide a better characterization of the systematics, which is the scope 
of the present work.

\subsection{Measuring the ICM physical parameters}

\begin{figure}
\includegraphics[width=0.49\textwidth]{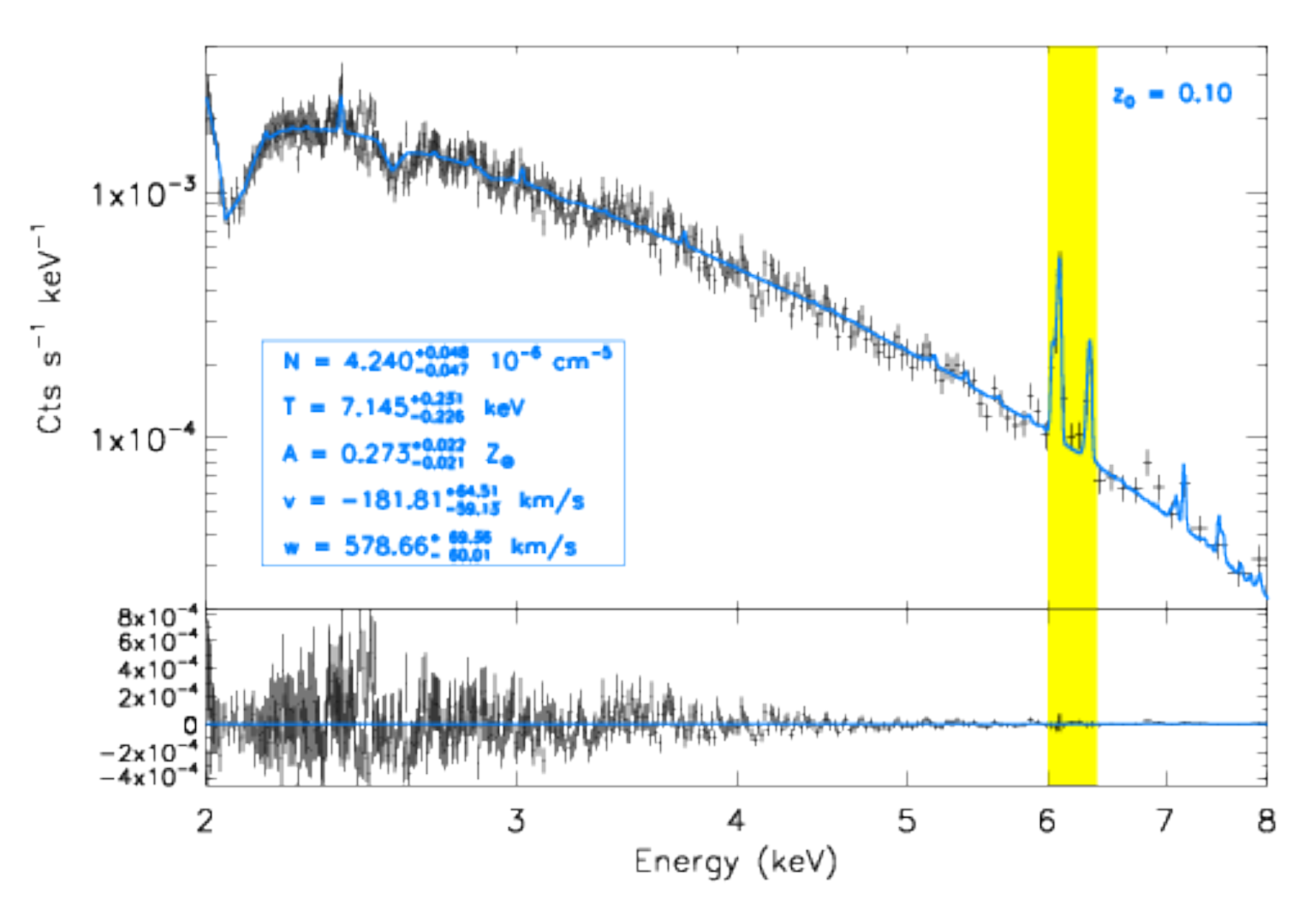}
\includegraphics[width=0.49\textwidth]{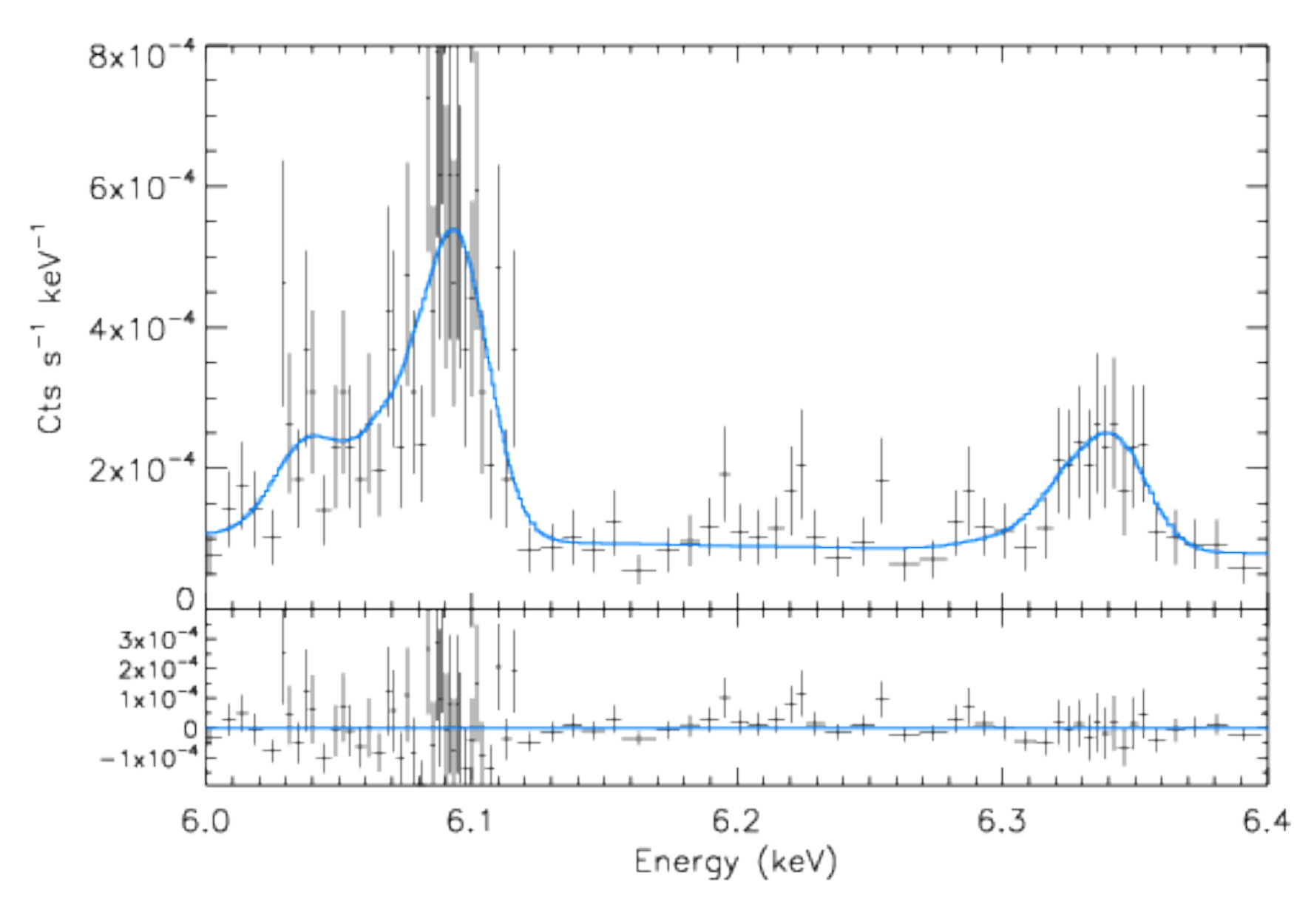}
\caption{
  Example of a mock X-IFU spectrum extracted from the \mn\,=\,0.75 run and of its fitting 
  procedure. Top panel: Spectral data points with error bars (black crosses) in the whole fitting 
  interval and best-fit model (blue line). Fit results, with errors, for the five free parameters 
  are indicated in the box in the bottom left. The value of $v$ is obtained assuming the true 
  redshift, $z_0=0.1$, is known. The bottom subpanel shows the residuals with respect to the 
  model. Bottom panel: Same as top panel, but zooming on the 6--6.4 keV energy range 
  (highlighted in yellow in the top panel) where the most prominent emission lines are 
  present. In this case the features correspond to the blending of the Fe~{\sc xxv} and 
  Fe~{\sc xxvi} K complexes. In both panels data points have been rebinned for 
  display purposes. The plot scaling for the spectra is logarithmic in the top panel and linear 
  in the bottom panel.
  }
\label{f:specfit}
\end{figure}

Our three mock event lists undergo a full X-ray analysis.  For each simulation 
we divide the X-IFU field of view using the Voronoi tessellation method \citep{cappellari03} 
in regions containing about 30000 photons each: this translates into about 600 pixels. Then we 
extract the spectra corresponding to each Voronoi region and, after binning them according
to the X-IFU response function, we fit them using \xspec\ (Cash statistics). To account for 
the effect of vignetting, we consider a different response function for each spectra: this 
is done by modifying the on-axis X-IFU response function according to the position of the 
center of the Voronoi region, following the same vignetting function as was implemented in 
\sixte. This step leads to a reduction in the effective area of 4\% to 8\%, depending on 
the energy, in the external regions of the FOV, enough to induce a significant systematic 
effect if not correctly accounted for. Instead, we neglect the effect of photon pile-up 
after verifying that it has a small impact in our final mock event lists. Once the response 
function is computed, we fit each spectra through the full 2--8 keV range using \xspec\ 
assuming a \bapec\ model with five free-parameters: normalization, 
temperature, metal abundance, redshift, and velocity broadening: this procedure is 
done ``blind'', i.e.,~without any a priori knowledge about the input physical 
conditions;  the only exceptions are the H column density, which we fix to the input value 
of $N_{\rm H}=5\times 10^{20}$ cm$^{-2}$, and  the value of $z_0$. We also point out 
that by adopting a single metallicity value, we are implicitly assuming that the abundance ratios of 
 the different ions (O/Fe, Si/Fe, and so on) have solar values. We
show in Fig.~\ref{f:specfit} a sketch of the fitting procedure and of its results applied 
to one of the \mn\,=\,0.75 mock spectra.

The fitting procedure proves to be robust and converges to a result without issues for 
the majority of the fitted spectra. In about 3\% of the cases an error occurs in the 
computation of the confidence regions;\footnote{The list of possible failures that may occur 
in \xspec\ during the error computation can be found in 
https://heasarc.gsfc.nasa.gov/xanadu/xspec/manual/XStclout.html} this is due both to the 
complexity of the fitting procedure (i.e.,~C-stat minimization on a five-parameters space) 
and to the intrinsic thermodynamical inhomogeneity of the emitting gas. To avoid possible 
issues we removed these cases from our analysis. 

\section{Results} \label{s:results}
\subsection{Intrinsic quantities definitions}

\begin{figure*}
\includegraphics[width=1.\textwidth]{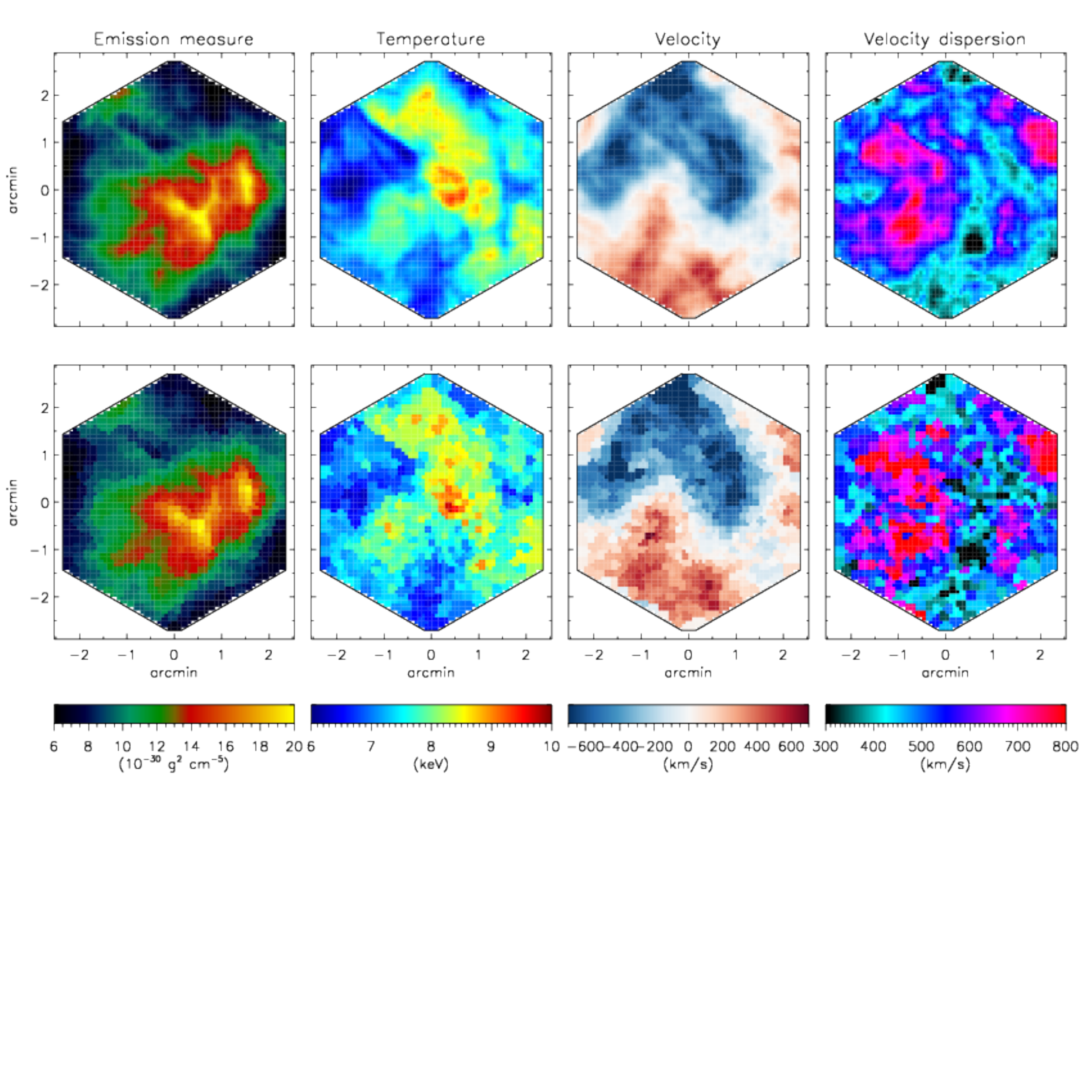}
\caption{
  X-IFU maps showing input quantities (top panels) vs measured ones (bottom) for the 
  \mn\,=\,0.75 model. Input quantities are derived directly from the simulation output, 
  considering the average along the line of sight enclosed by the X-IFU pixel 
  ($4.38$ arcsec wide) and computed with the definitions in 
  Eqs.(\ref{e:em})--(\ref{e:wew}). For measured quantities we show the nominal value 
  derived from the fitting of our synthetic spectra (see text for details) in the 
  corresponding 588 Voronoi regions ($\approx 125$ arcsec\sqr\ each). From left to 
  right: emission measure, gas temperature (spectroscopic-like, in the top panel), 
  velocity, and velocity dispersion (both emission-weighted, in the top panels).
  }
\label{f:maps}
\end{figure*}

We show in Fig.~\ref{f:maps} (bottom panels) the X-IFU maps with the results of our fitting 
procedure for the \mn\,=\,0.75 run. Measured velocity maps were obtained from the redshift 
nominal value by reverting Eq.~\ref{e:zs}. We compared these maps with those of 
the corresponding projected quantities computed directly from the simulations (top panels), 
averaging in the volume enclosed in the same line of sight. From the various possible 
definitions of average (i.e.,~mass-weighted, emission weighted), we chose the ones that show a 
better correlation with the measured values. In detail for each X-IFU pixel we compute the following 
quantities:
\begin{enumerate}
\item emission measure, defined as
\begin{equation}
{\rm EM} \equiv \int \rho^2 \, dl 
\ \ \longrightarrow \ \
\sum_i \rho_i^2 \, l_{\rm pix} \, ;
\label{e:em}
\end{equation}
\item spectroscopic-like temperature \citep{mazzotta04},
\begin{equation}
T_{\rm sl} \equiv \frac{\int \rho^2 T^{0.25} \, dl}{\int \rho^2 T^{-0.75} \, dl}
\ \ \longrightarrow \ \
\frac{\sum_i \rho_i^2 T_i^{0.25}}{\sum_i \rho_i^2 T_i^{-0.75}} \, ;
\label{e:tsl}
\end{equation}
\item emission-weighted velocity
\begin{equation}
v_{\rm ew} \equiv \frac{\int \rho^2 v\, dl}{\int \rho^2 \, dl}
\ \ \longrightarrow \ \
\frac{\sum_i \rho_i^2 v_i}{\sum_i \rho_i^2} \, ;
\label{e:vew}
\end{equation}
\item and emission-weighted velocity dispersion
\begin{equation}
w_{\rm ew}^2 \equiv \frac{\int \rho^2 v^2\, dl}{\int \rho^2 \, dl} - v_{\rm ew}^2
\ \ \longrightarrow \ \
\frac{\sum_i \rho_i^2 v_i^2}{\sum_i \rho_i^2} - v_{\rm ew}^2 \, .
\label{e:wew}
\end{equation}
\end{enumerate}
In the above formulas on the left-hand side we show the analytical definitions with  
line-of-sight integrals, and on the right-hand side the equivalent numerical definitions with the index $i$ running 
through all the cells along the full length of the 
simulation box ($\simeq$\,1.4\,Mpc). While the definition of EM (directly connected to the spectrum 
normalization, as in Eq.~\ref{e:norm}) and $T_{\rm sl}$ are already widely 
adopted in X-ray astrophysics, only recently some works \citep[see, e.g.,][]{biffi13} have proved 
how the last two quantities correlate with the centroid-shift and line broadening measurements 
better than the equivalent mass-weighted values. A visual inspection shows an excellent agreement 
of both emission measure and velocity. In particular, it is important to note that with 
our mock measurements we are able to recover the details of the line-of-sight velocity field 
with great accuracy. In addition, measured temperature maps show a good agreement with spectroscopic-like 
temperature maps. Broadening measurements, on the other hand, prove to be the most challenging 
even though both the average broadening value and the main 
features of the \wew\ maps are appropriately recovered.

\subsection{Biases and uncertainties} \label{ss:bias}

In this section we analyze the accuracy of the X-ray measurements by comparing them with the 
input quantities computed from the simulations with the definitions of Eqs.~\ref{e:em}--\ref{e:wew}. 
In the following, we will distinguish between biases, i.e., possible systematic offsets in the 
measurements, and systematic errors/uncertainties, i.e., the irreducible scatter between 
measurements and the corresponding theoretical values. For the analysis presented here, the latter values 
are  associated with the limits of the assumption of a five-parameter model adopted in our fitting 
procedure in the case of ICM mixing. These errors may eventually be reduced by adopting more 
complicated models (such as allowing multiple ICM components).

\begin{figure*}
\includegraphics[width=1.\textwidth]{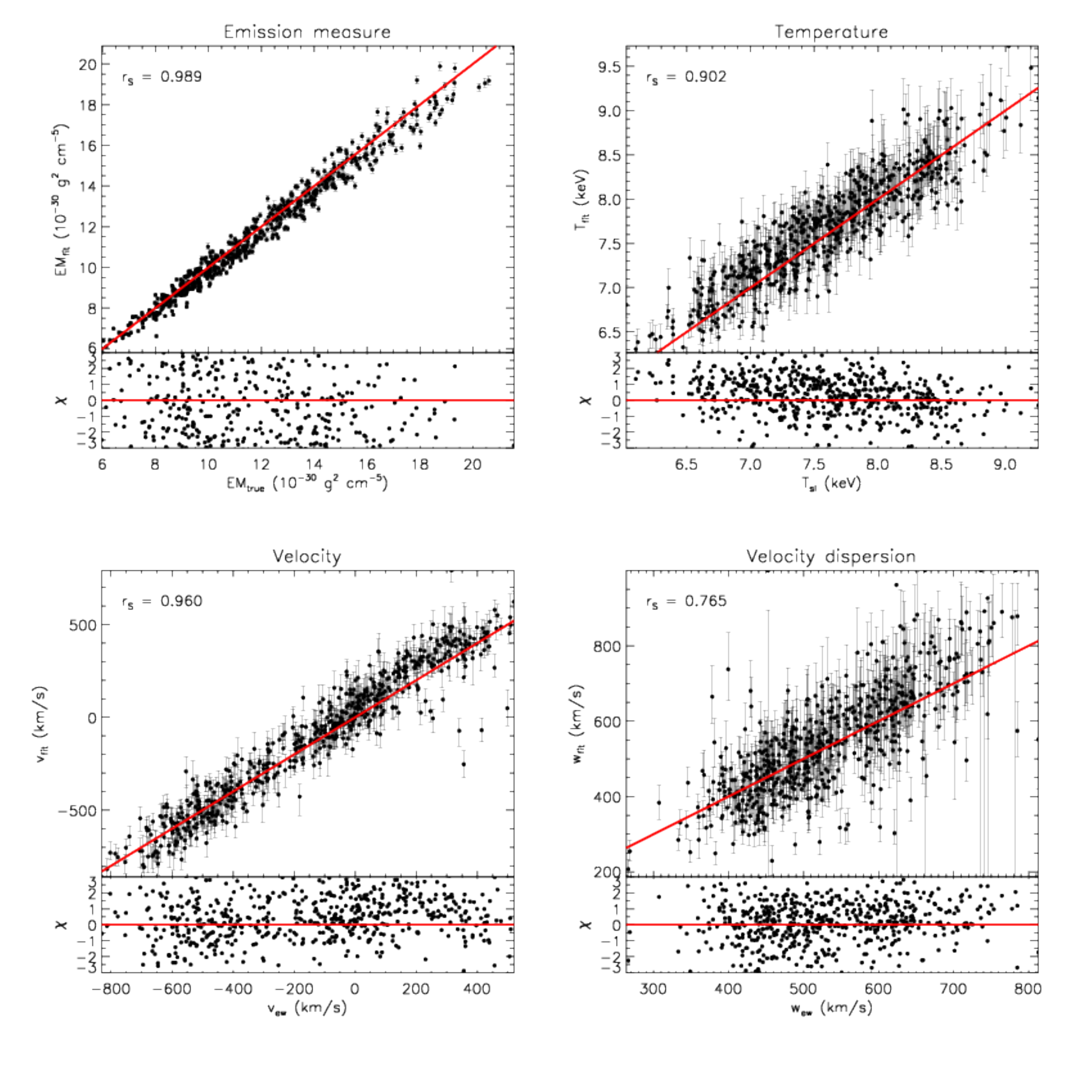}
\caption{
  Scatter plots of observed quantities with errors ($y$-axis)  vs input projected values ($x$-axis) 
  for the \mn\,=\,0.75 simulation. Top left: Emission measure. Top right: Temperature vs 
  spectroscopic-like temperature. Bottom left: Velocity vs emission-weighted velocity. 
  Bottom right: Velocity dispersion vs emission-weighted velocity dispersion. In each 
  plot the red solid lines shows the identity and the bottom panel shows the residuals in 
  units of $\sigma$. In the top left corner of each panel we show the Spearman correlation rank 
  between observed and input values.
  }
\label{f:scp}
\end{figure*}

We show in Fig.~\ref{f:scp} the relation among the results obtained from the fitting 
procedure, with errors, and those extracted from the simulations for the \mn\,=\,0.75 model (the 
red solid lines indicate the identity). In Appendix~\ref{a:scplots} we show the same plots for the 
other two simulations. The values on the $x$-axes were computed by applying the 
formulas of Eqs.~\ref{e:em}--\ref{e:wew} directly to the input hydrodynamical simulation in the volume 
corresponding to the Voronoi region of the measurements projected along the line of sight.
As expected from Fig.~\ref{f:maps}, emission measure and velocity show the most 
precise measurements. In particular, we point out that the comparison between the fit result and 
\vew\ shows no significant bias even for high values of  velocity. Temperature measurements 
show also a good agreement with \tsl, with a small hint of temperature overestimation for the few 
regions with \tsl$<6.5$ keV; this may indicate that in the lower temperature regimes the definition 
of \tsl\ provided in Eq.~\ref{e:tsl}, which was optimized for CCDs, might need to be revisited. As 
expected,  the velocity dispersion measurements also show good agreement with the corresponding \wew, with 
possibly a hint of overestimation for high \wew\ values. We note 
that in this case errors in the measurements may vary considerably from $\sigma_w 
\approx 40$ km\,s$^{-1}$ to $\sigma_w \approx 100$ km\,s$^{-1}$. In the top left corner of each plot we 
show the Spearman correlation rank between measured and real quantities. In all cases 
the correlation is extremely high. Indeed, we obtain an almost perfect correlation ($r_S \approx 
1$) for emission measure and velocity, to $r_S \sim 0.9$ for temperature, and $r_S \sim 
0.75$ for velocity dispersion. This confirms the accuracy of the measurements of the different 
physical quantities.

\begin{table*}
\begin{center}
\caption{
  Statistical estimates for the accuracy on the measurements of the five physical quantities compared to the input 
  values in the three simulations. Column 1: simulation model (Mach number). Column 2: average value of the quantity 
  in the simulated map. 
  Column 3: average bias in the measurement (see Eq.~\ref{e:bias}). Columns 4--6: total, statistical and 
  systematic errors (Eqs.~\ref{e:stot}--\ref{e:ssys}). Except for velocity, we also quote in parentheses the values as a percentage of 
  the average (Col. 2). Columns 7 and 8: fraction of outliers in the $\chi$ distribution 
  ($2.5\sigma$ clipping) and fraction of measurements within 1$\sigma$ with respect to the reference value, respectively.
  }
\begin{tabular}{rccccccc}
\hline
\hline
\noalign{\vskip 0.1cm}
 & \multicolumn{7}{c}{Emission measure ($10^{-30}$ g$^2$/cm$^5$)} \\
Model & $\langle {\rm EM} \rangle$ & $b_{\rm EM}$  & $\sigma_{{\rm EM},\rm tot}$ & $\sigma_{{\rm EM},\rm stat}$ & $\sigma_{{\rm EM},\rm sys}$ & $f_{\rm out}$ & $\chi^2<1$ \\
 \mn = 0.25 &  32.71 &  -0.30 (-0.84\%) &     1.02 (3.12\%) &     0.34 (1.04\%) &     0.96 (2.94\%) & 4.6\% & 21.7\% \\ 
 \mn = 0.50 &  19.53 &  -0.21 (-1.01\%) &     0.70 (3.60\%) &     0.21 (1.09\%) &     0.67 (3.43\%) & 1.2\% & 20.3\% \\ 
 \mn = 0.75 &  11.86 &  -0.19 (-1.46\%) &     0.51 (4.31\%) &     0.13 (1.12\%) &     0.49 (4.16\%) & 1.5\% & 18.4\% \\ 
 \hline
\noalign{\vskip 0.1cm}
 & \multicolumn{7}{c}{Temperature (K)} \\
Model & $\langle T_{\rm sl} \rangle$ & $b_T$  & $\sigma_{T,\rm tot}$ & $\sigma_{T,\rm stat}$ & $\sigma_{T,\rm sys}$ & $f_{\rm out}$ & $\chi^2<1$ \\
 \mn = 0.25 &   8.18 &   0.03 (0.42\%) &     0.28 (3.38\%) &     0.26 (3.14\%) &     0.10 (1.24\%) & 1.7\% & 62.8\% \\ 
 \mn = 0.50 &   7.59 &   0.05 (0.72\%) &     0.25 (3.24\%) &     0.22 (2.92\%) &     0.11 (1.39\%) & 0.7\% & 60.2\% \\ 
 \mn = 0.75 &   7.57 &   0.08 (1.09\%) &     0.29 (3.79\%) &     0.24 (3.11\%) &     0.16 (2.16\%) & 2.4\% & 56.6\% \\ 
 \hline
\noalign{\vskip 0.1cm}
 & \multicolumn{7}{c}{Metal abundance (\zsun)} \\
Model & $\langle Z \rangle$ & $b_Z$  & $\sigma_{Z,\rm tot}$ & $\sigma_{Z,\rm stat}$ & $\sigma_{Z,\rm sys}$ & $f_{\rm out}$ & $\chi^2<1$ \\
 \mn = 0.25 &   0.30 &  -0.00 (-0.25\%) &     0.02 (7.41\%) &     0.02 (6.95\%) &     0.01 (2.57\%) & 1.5\% & 65.2\% \\ 
 \mn = 0.50 &   0.30 &  -0.01 (-2.09\%) &     0.02 (7.90\%) &     0.02 (6.93\%) &     0.01 (3.78\%) & 0.9\% & 60.2\% \\ 
 \mn = 0.75 &   0.30 &  -0.01 (-3.36\%) &     0.02 (8.31\%) &     0.02 (7.29\%) &     0.01 (3.99\%) & 0.3\% & 60.2\% \\ 
 \hline
\noalign{\vskip 0.1cm}
 & \multicolumn{7}{c}{Velocity (km\,s$^{-1}$)} \\
Model & $\langle v_{\rm ew} \rangle$ & $b_v$  & $\sigma_{v,\rm tot}$ & $\sigma_{v,\rm stat}$ & $\sigma_{v,\rm sys}$ & $f_{\rm out}$ & $\chi^2<1$ \\
 \mn = 0.25 &  -16.05 &   2.73  &    34.21 &    24.84 &    23.51 & 4.6\% & 52.2\% \\ 
 \mn = 0.50 & -127.63 &  -8.61  &    66.24 &    41.75 &    51.43 & 6.1\% & 43.9\% \\ 
 \mn = 0.75 & -132.43 &  24.51  &   103.08 &    66.66 &    78.62 & 3.8\% & 48.5\% \\ 
 \hline
\noalign{\vskip 0.1cm}
 & \multicolumn{7}{c}{Velocity dispersion (km\,s$^{-1}$)} \\
Model & $\langle w_{\rm ew} \rangle$ & $b_w$  & $\sigma_{w,\rm tot}$ & $\sigma_{w,\rm stat}$ & $\sigma_{w,\rm sys}$ & $f_{\rm out}$ & $\chi^2<1$ \\
 \mn = 0.25 & 177.82 &  10.10 (5.38\%) &    31.49 (17.71\%) &    22.16 (12.46\%) &    22.37 (12.58\%) & 1.5\% & 52.0\% \\ 
 \mn = 0.50 & 340.77 &   3.73 (1.29\%) &    65.13 (19.11\%) &    37.60 (11.03\%) &    53.18 (15.61\%) & 5.2\% & 45.3\% \\ 
 \mn = 0.75 & 532.08 &  11.73 (1.94\%) &    94.12 (17.69\%) &    64.18 (12.06\%) &    68.84 (12.94\%) & 4.3\% & 52.7\% \\
\hline
\hline
\label{t:res}
\end{tabular}
\end{center}
\end{table*}

The quantification of these results is shown in Table~\ref{t:res} together with other statistical estimates 
described and discussed in the following. All results are reported for the three different simulation setups, 
and are also related to the metallicity. We define the average bias (third column) as 
\begin{equation}
b \equiv \langle m_i-t_i \rangle \, ,
\label{e:bias}
\end{equation}
where $m_i$ and $t_i$ are the sets of measured and true projected quantities extracted from the simulations 
(Eqs.~\ref{e:em}--\ref{e:wew}).  Remarkably, biases in the measurements are almost negligible 
for all quantities: this is both an indication of the robustness of the fitting procedure and of the suitability of 
the definitions of Eqs.~\ref{e:em}--\ref{e:wew} in representing the observationally derived values. On the other 
hand, we verified that mass-weighted quantities are, instead, significantly different with respect to measured 
ones, with mass-weighted temperature and velocity dispersion biased high by $\sim$0.2 keV and $\sim$40 
km\,s$^{-1}$, respectively. Mass weighted velocities are instead about 30\% smaller than observed ones (see also 
\citealt{biffi13}).

In the central columns of the table, we quote the total, statistical, and systematic errors  of the 
measured quantities computed in the following way. Total errors are computed as
\begin{equation}
\sigma_{\rm tot}^2=\langle (m_i-t_i)^2 \rangle \, .
\label{e:stot}
\end{equation}
Statistical errors are instead derived from the fit errors with the following formula (see details in 
Appendix~\ref{a:sstat}):
\begin{equation}
\sigma_{\rm stat}^2=\left\langle \left(\sigma_2-\sigma_1\right)^2 + \sigma_1 \sigma_2 \right\rangle \, .
\label{e:sstat}
\end{equation}
This expression is also appropriate  for asymmetric errors, with $\sigma_1$ and $\sigma_2$ being the left and right 
errors, respectively. Thus, in our case it provides a better description with respect to the simple Gaussian error 
estimates. Systematic errors are then simply derived as
\begin{equation}
\sigma_{\rm sys}^2 = \sigma_{\rm tot}^2 - \sigma_{\rm stat}^2 \, ,
\label{e:ssys}
\end{equation}
and provide an indication of the expected scatter associated with the model uncertainty.

For emission measure, the number counts adopted in our simulations are enough 
to reduce statistical errors below systematic ones ($\sim 3$\%). The latter are associated with the X-IFU point 
spread function that scatters photons in the areas close to the borders of the Voronoi regions, and would be 
negligible with a more  realistic exposure time. This  also causes the highest EM measurements to be slightly 
underestimated, as it appears in Fig.~\ref{f:scp}.
Most interestingly, we observe that the values of $\sigma_{v, \rm sys}$ increase for 
larger Mach numbers, going from 25 km\,s$^{-1}$ for \mn\,=\,0.25 to about 80 km\,s$^{-1}$ for \mn\,=\,0.75, due to the larger 
gas mixing. The same applies for $\sigma_{w,\rm sys}$, which grows proportionally with $\langle w_{\rm ew} \rangle$ 
and with the Mach number, and remains approximately 15\% of the average value.

In order to evaluate the accuracy of the errors provided by our fitting procedure, for each measurement 
we compute its normalized residual
\begin{equation}
\chi \equiv \frac{m-t}{\sigma_{1,2}} \, ,
\end{equation}
being $\sigma_{1,2}$ the left and right measurement errors provided by the fit. We then use the distributions of the 
$\chi$ values to determine two statistical estimators. First, we identify the outliers by applying a 2.5$\sigma$ clipping to 
the different distributions and compute its fraction. Then we compute the fraction of values in the range $[-1,1]$: this represents the 
fraction of points within $1 \sigma$ of the true value. The results are listed in the last two columns of Table~\ref{t:res}. 
The outliers fraction for the measurements of EM, $T$, and $Z$ are comparable to expected values for the perfectly 
Gaussian case (1.25\%)\footnote{We verified that the high number of outliers in the EM measurements for the \mn\,=\,0.25 
simulation is due to a single region where a shock is occurring.}. On the other hand, measurement of $v$ and $w$ show a 
significant number of outliers: this can happen in regions with large gas mixing when a single velocity value is not able to 
describe the properties of the emission lines.

The values shown in the last column indicate that our fitting procedure is underestimating the true dispersion of the values of $\chi$ 
(all values are smaller than 68.8\%, expected for perfectly estimated Gaussian errors). Most notably, this figure is about 45--50\% for $v$ 
and $w$ due both to the presence of outliers and systematic errors. For $T$ and $Z$ measurements, instead, where statistical errors 
dominate, values are close the expected Gaussian behavior with a fraction of measurements within 1$\sigma$ of $\sim$60\%.

\subsection{Cross-correlation among measured and input values} \label{ss:ccor}

\begin{figure*}
  \centering
  \begin{minipage}[b]{0.49\textwidth}
    \includegraphics[width=\textwidth]{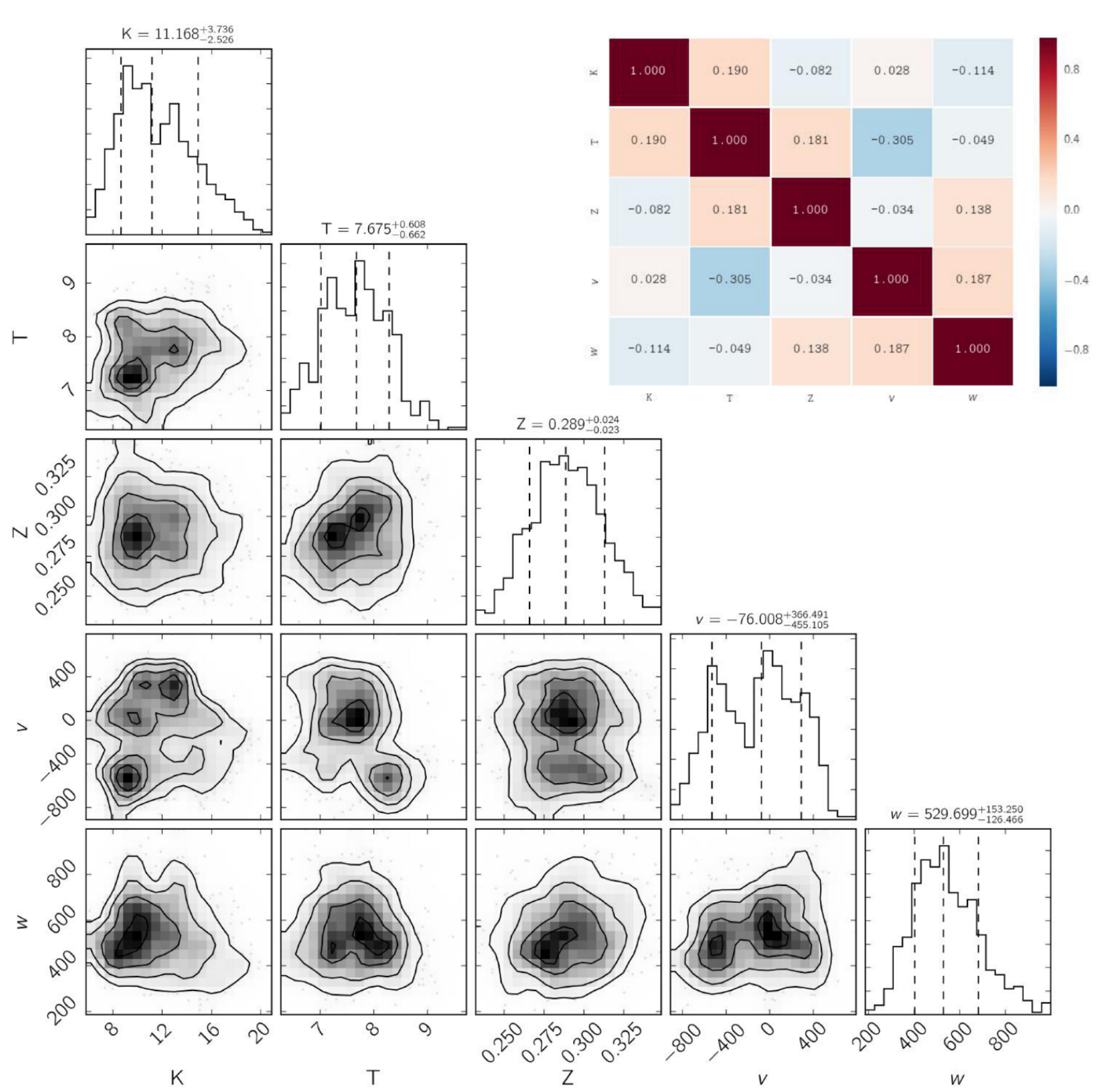}
  \end{minipage}
  \hfill
  \begin{minipage}[b]{0.49\textwidth}
    \includegraphics[width=\textwidth]{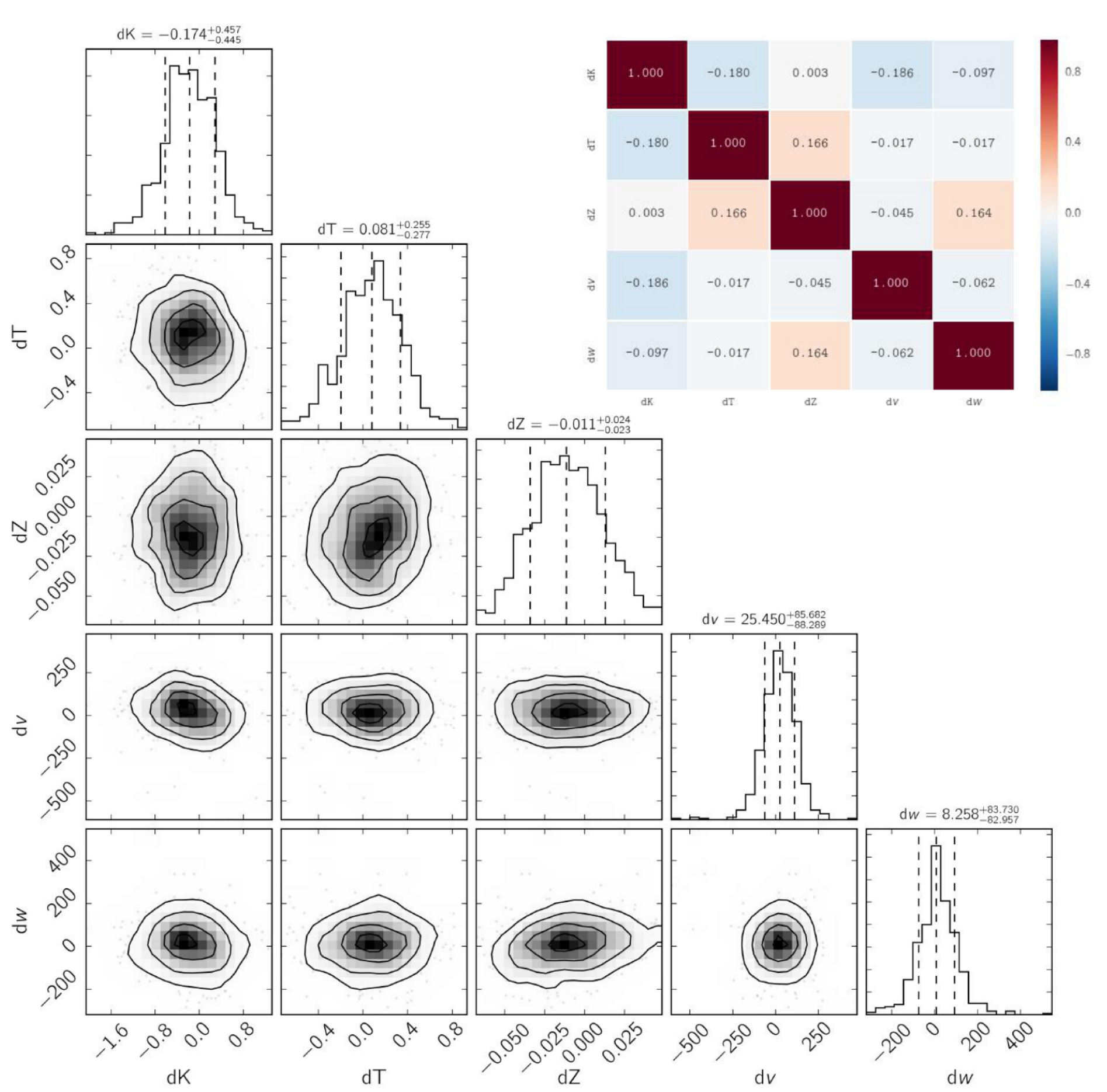}
  \end{minipage}
  \caption{Left panel: Corner plot and heat map for the measured best-fit spectral quantities in the 
  case of \mn\,=\,0.75. The distributions (median, 1st and 3rd quartiles) and the correlations among the 
  measured quantities are shown. The heat map is color-coded and indicates the level of correlations among 
  the same quantities. Right panel: Same as left panel, but evaluated using the differences between 
  measured and reference values.}
 \label{f:crn}
\end{figure*}

We have investigated the level of cross-correlation among both the measured quantities and the differences between 
measured and input values (Fig.~\ref{f:crn}, left  and right panel, respectively).
The only significant deviations from the null-hypothesis of no correlation (with $|\rho|>0.2$, 
corresponding to a P-value $< 10^{-6}$ for the dataset investigated)
is between $T$ and velocity $v$ ($\rho=-0.31$). This correlation is induced by the bulk motion of some hot gas
at about $-500$ km\,s$^{-1}$, which is well resolved in all the plots of the velocity.

The lack of any correlation between the spectroscopic measurements and the estimated
differences between the same measurements and the input values guarantees the robustness of the
outputs of the spectral fitting procedure, in particular against any possible degeneracy in the estimates
of the best-fit parameters. This is a non-trivial result because the same emission lines are used to constrain 
(i) the metal abundance through the equivalent width of the line against the continuum that defines the gas temperature,
(ii) the velocity $v$ from the centroid of the line, (iii) the velocity dispersion $w$ from its broadening.
Thus, we can conclude that these measurements are independent  from each other at high statistical significance.

\section{Structure functions} \label{s:sf}

Most of the information on the kinematics of the ICM can be derived from the analysis of the 
fluctuations of the velocity and density field. Ideally, the injection/dissipation 
scale and inertial range should be physically constrained by the 3D power spectrum (e.g.,~\citealt{Schuecker:2004,churazov12,zhuravleva12,Gaspari:2013_coma}); however, it is still possible to 
derive meaningful pieces of information from their 2D counterparts \citep[e.g.,][]{zuhone16}. In this work, we only focus  on studying the 
2D fluctuations of surface brightness, velocity, and velocity dispersion of our simulation set and estimate the accuracy of future X-IFU 
observations to derive their 2D power spectra.

Following \cite{zuhone16}, we analyze the structure functions of our X-IFU maps for the quantities of 
interest, and compare them with those derived from the simulations. For a given physical distance 
$r$, the second-order structure function of a given mapped quantity, $m(\vec x)$, is defined as
\begin{equation}
{\rm SF}(r) \equiv \langle |m(\vec x + \vec r) - m(\vec x)|^2 \rangle \, .
\label{e:sf}
\end{equation}
The structure function is mathematically related to the autocorrelation function and to the 2D 
power spectrum, but proves to be particularly useful in our case since it allows us to determine 
the bias associated with statistical and systematic errors and remove it, provided that these 
uncertainties are modeled correctly. Specifically, when considering the squared differences of 
a set of measured quantities in the same distance bin, including errors in the expression we 
can write
\begin{equation}
s_{ij}' = [(m_i + \delta m_i) - (m_j +\delta m_j)]^2 \, 
,\end{equation}
being $\delta m_i$ and $\delta m_j$ the differences between the measured quantities and the 
true ones $m_i$ and $m_j$, respectively. By expanding the expression and averaging in the case of symmetric 
Gaussian errors, $\sigma_{\rm tot}$, it can be shown that \citep[see Appendix C in][]{zuhone16}
\begin{equation}
\langle s_{ij} \rangle_r = \langle s_{ij}' \rangle_r - 2\sigma_{\rm tot}^2 \, ,
\label{e:sij}
\end{equation}
where the symbol $\langle {\rm ...} \rangle_r $ indicates the average between 
pairs at the same distance $r$. The previous expression shows that since both $v$ and $w$ show an 
intrinsic variability comparable 
to that associated with measurement errors (see Table~\ref{t:res}), a good model of the latter 
is necessary to achieve a precise determination of their structure function. However, it should be kept in mind that in real observations $\sigma_{\rm tot}$ is not known since it also includes  
the systematics that are unknown to the observer. To mimic the observational approach we  compute the structure 
function of our observed X-IFU maps by modeling only the effect of statistical errors, which we can 
estimate from the errors in the fit results. Therefore, we modify Eq.~\ref{e:sij} as
\begin{equation}
\langle s_{ij} \rangle_r = \langle s_{ij}' \rangle_r - 2 \langle \sigma_{{\rm stat},ij}^2 \rangle_r \, ,
\label{e:sfp}
\end{equation}
where the last term represents the average of the statistical variance of the set of points at the 
same distance $r$. Since in our results we have non-symmetric errors, each value of 
$\sigma_{{\rm stat},ij}^2$ is derived from the left-hand and right-hand error as in Eq.~\ref{e:sstat}.

Considering all of the above, we derive the best observational estimate of the (2D) structure 
functions of surface brightness fluctuations $\delta$, $v$, and $w$, respectively, with the following 
formulas:
\begin{equation}
{\rm SF}_{\delta} (r) = \langle |\delta(\vec x + \vec r) - \delta(\vec x)|^2 - 2 \sigma_{\delta, \rm stat}^2 \rangle \, ,
\label{e:sfd}
\end{equation}
\begin{equation}
{\rm SF}_v (r) = \langle |v(\vec x + \vec r) - v(\vec x)|^2 - 2 \sigma_{v, \rm stat}^2 \rangle \,
\label{e:sfv}
,\end{equation}
and
\begin{equation}
{\rm SF}_w (r) = \langle |w(\vec x + \vec r) - w(\vec x)|^2 - 2 \sigma_{w, \rm stat}^2 \rangle \, .
\label{e:sfw}
\end{equation}
The quantity $\delta$ that appears in Eq.~\ref{e:sfd} is defined as
\begin{equation}
\delta(\vec x) \equiv \frac{{\rm EM}(\vec x)}{\langle {\rm EM}(r) \rangle}-1 \, .
\label{e:delta}
\end{equation}
The corresponding statistical error $\sigma_{\delta, \rm stat}=\sigma_{{\rm EM}, \rm stat}/\langle {\rm EM}(r) 
\rangle$, being $\langle {\rm EM}(r) \rangle$ the average surface brightness of the points at the same distance $r$ 
from the peak of the emission measure map; this conversion allows us to study the surface brightness fluctuations 
with respect to the average profile. We note that the SF$_v$ (at large $r$) is tied to the variance of the 
turbulence field (i.e.,~specific kinetic energy), while the SF$_w$ (line broadening) is linked to the variance of 
the turbulent velocity, which can be interpreted as a measure of turbulence intermittency.

\begin{figure}
\includegraphics[width=0.49\textwidth]{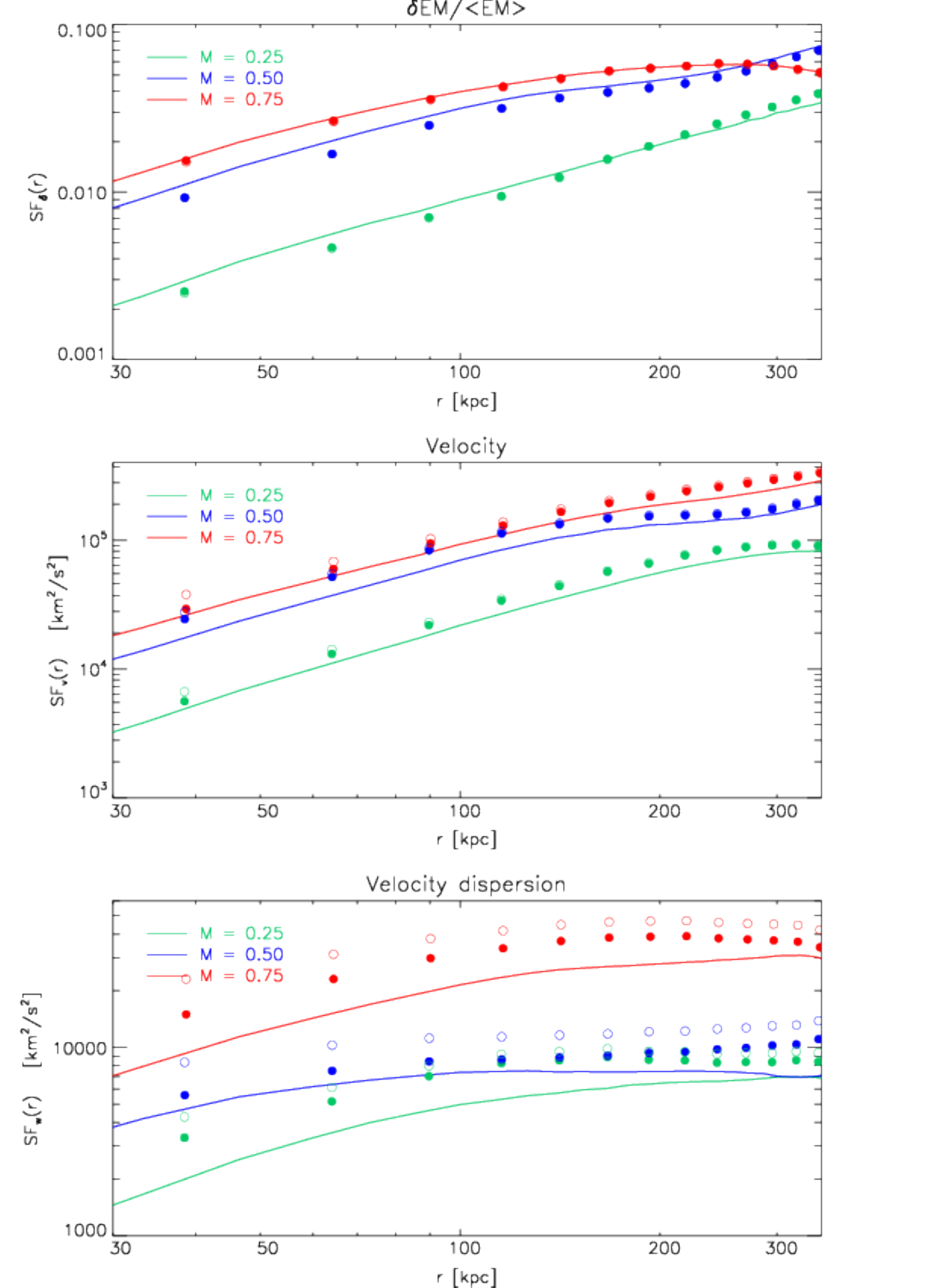}
\caption{
  Second-order 2D structure functions of emission measure fluctuations (top), velocity 
  (middle), and velocity dispersion (bottom) as a function of distance for the three \mnall\ 
  simulations (green, blue, and red, respectively). In each plot the measured SFs derived from our 
  simulated X-IFU fields (filled dots with error bars) is compared to those of $\delta$, \vew,\ and \wew\ 
  (see  definitions in Eqs. \ref{e:em}, \ref{e:vew}, and \ref{e:wew}) derived directly from the 
  hydrodynamical simulations in the same field of view (solid lines). Error bars represent the 
  error on the mean computed with 100 bootstrap samplings (in most cases  smaller than 
  the dot size). Open circles indicate the values of the measured SFs without the statistical 
  error subtraction (see Eqs. \ref{e:sfv} and \ref{e:sfw}). The FOV covers only a 
  small fraction of the cluster, namely the core region.
  }
\label{f:stf}
\end{figure}

It is important to note that the region covered by our X-IFU simulations, i.e.,~the core of our simulated cluster, allows us to estimate fluctuations up to a maximum distance of 350 kpc. In order to obtain a complete representation of the SFs over the full cluster volume (fully capturing the injection scale) we would need to enlarge the total area by a factor of $\sim$10, with multiple X-IFU pointings, which is beyond the scope of this work. In addition, the SFs measured in the core can significantly vary from the global SFs (which describe the full physical cascade\footnote{We verified that taking multiple projections and a large cube reproduces the proper Kolmogorov 2D SF$_v$ with $5/3$ slope and gradual flattening only at several 100 kpc near the injection scale.}) because of the cosmic variance and related stochastic nature of the turbulent eddies along a given line of sight. As a reference, we verified that the logarithmic scatter in the amplitude of the SFs due to the different line-of-sight projection in our simulations is $\Delta \, A_{\rm Log}\simeq 0.10-0.15$ (at $r<100$ kpc).

The comparison between observationally derived and input SF$_\delta$, SF$_v$, and SF$_w$ is shown in Fig.~\ref{f:stf}. To understand the 
effect of the statistical error correction, we also show (open circles) the same results neglecting 
the $- 2 \sigma_{\rm stat}^2$ terms. The structure function of the emission measure fluctuations $\delta$ 
is recovered almost perfectly, 
with only a minor difference for the \mn\,=\,0.5 simulation. This is expected given the high correlation 
between measured and input quantities (see the discussion in Section \ref{ss:bias}) 
and the small measurement errors. The general shape of the structure function of the velocity field is 
also recovered, with the three physical models clearly distinguishable. However, measured values show a 
non-negligible overestimate (about 5--10\%); this can be partially explained as systematic errors that are 
not accounted for in our computation and that would introduce a $- 2 \sigma_{v, \rm sys}^2$ term in the 
equation. We also point out that the modeling of error subtraction described by Eq.~\ref{e:sfp} 
works under the hypothesis of both uniform and Gaussian errors, which is not completely true due to the presence
of a non-negligible number of outliers, as 
discussed in Section~\ref{ss:bias}. This means that in order to achieve a more robust characterization 
of SF$_v$, a more accurate error model is required. All these considerations also apply  to SF$_w$, with 
a significantly larger overestimate associated with the error subtraction.

We analyze the impact of these systematics in determining the properties of the 2D 
power spectrum, \ptd, of the different quantities. As pointed out in \cite{zuhone16}, 
if we assume 
a power-law form of the power spectrum in the inertial range (i.e.,~below the injection scale), \ptd$(k) \propto k^\alpha$, then the 
corresponding SF scales as $r^\gamma$, with $\gamma=-(\alpha+2)$. Therefore, we fit 
the SFs of Fig.~\ref{f:stf} using 
\begin{equation}
{\rm SF}(r) = A \left(\frac{r}{r_0}\right)^\gamma \, ,
\label{e:plaw}
\end{equation}
with $r_0$ fixed to 100 kpc, while $A$ and $\gamma$ are left as free parameters. 
We compute the fit in the interval 30--120 kpc to ensure that in all cases we have functions that 
increase with $r$ and avoid the flattening due to the large-scale plateau. 
Finally, by comparing the resulting best-fit values obtained from our mock measurements with those 
obtained directly from the projected simulation quantities, we are able to estimate the impact of the 
systematics in the structure function on the slope and normalization of the SF itself and, consequently, of 
their \ptd.

\begin{table}
\begin{center}
\caption{
 From top to bottom: Normalization and slopes of the structure functions, SF$(r)$, of emission 
 measure fluctuations, velocity, and velocity dispersion, for the three simulations with \mnall. Column 1: 
 simulation model (Mach number). Columns 2 and 3: normalization derived from simulations and from measured quantities, 
 respectively. Units are shown in parenthesis. Columns 4 and 5: same as Cols. 2 and 3, respectively, but for 
 the slope. The fitting range is 30--120 kpc. 
}
\begin{tabular}{rcccccc}
\hline
\hline
\noalign{\vskip 0.1cm}
 && \multicolumn{5}{c}{Emission measure fluctuations ($10^{-3}$)} \\
 Model &&  $A_{\delta,{\rm sim}}$ & $A_{\delta,{\rm meas}}$ && $\gamma_{\delta,{\rm sim}}$ & $\gamma_{\delta,{\rm meas}}$  \\
 \mn = 0.25 &&  9.15 &  8.00 &&  1.17 &  1.20\\
 \mn = 0.50 && 31.85 & 27.56 &&  1.09 &  1.13\\
 \mn = 0.75 && 40.29 & 38.57 &&  0.96 &  0.93\\
 \hline
\noalign{\vskip 0.1cm}
 && \multicolumn{5}{c}{Velocity ($10^{ 4}$ km$^2$/s$^2$)} \\
 Model &&  $A_{v,{\rm sim}}$ & $A_{v,{\rm meas}}$ && $\gamma_{v,{\rm sim}}$ & $\gamma_{v,{\rm meas}}$  \\
 \mn = 0.25 &&  2.19 &  2.66 &&  1.56 &  1.63\\
 \mn = 0.50 &&  7.00 &  9.49 &&  1.45 &  1.41\\
 \mn = 0.75 &&  9.32 & 10.76 &&  1.33 &  1.36\\
 \hline
\noalign{\vskip 0.1cm}
 && \multicolumn{5}{c}{Velocity dispersion ($10^{ 4}$ km$^2$/s$^2$)} \\
 Model &&  $A_{w,{\rm sim}}$ & $A_{w,{\rm meas}}$ && $\gamma_{w,{\rm sim}}$ & $\gamma_{w,{\rm meas}}$  \\
 \mn = 0.25 &&  0.40 &  0.75 &&  0.94 &  0.84\\
 \mn = 0.50 &&  0.59 &  0.85 &&  0.46 &  0.41\\
 \mn = 0.75 &&  1.72 &  3.12 &&  0.87 &  0.75\\
 \hline
 \hline
\label{t:stffit}
\end{tabular}
\end{center}
\end{table}

We show the results of the fit in Table~\ref{t:stffit}. It is clear that the only significant impact of the systematics in our SFs is in the normalization estimate. The value of 
$A_{v,{\rm meas}}$ exceeds the input value by 15--30\%, while for $A_{w,{\rm meas}}$ it can be almost doubled. On the other hand, the slopes of the initial structure functions are correctly recovered in all cases, with maximum differences between input and measured values of $\sim 0.1$.\footnote{The SF$_v$ slopes for the \mn\,=\,0.5 and \mn\,=\,0.75 runs are somewhat smaller than the Kolmogorov slope ($\gamma=5/3$), which characterizes the subsonic turbulence stirring the whole cluster. We verified that this is due to the variance associated with the limited FOV.}
This confirms that the systematics in the SFs measurements are induced by an underestimation of the $-2\sigma_{\rm stat}^2$, which boosts our measurements but has a minor impact on the slope.


\section{Summary and conclusions} \label{s:concl}

We have developed, for the first time, an end-to-end mock X-ray analysis pipeline that mimics future 
\emph{Athena} X-IFU observations. Our goal was to quantify possible systematics that may arise in the 
measurements of the main projected physical properties of the ICM, emission measure, temperature, and 
metallicity;  most notably, the two key quantities that 
will be made available via high-resolution spectroscopic imaging are the line-of-sight velocity $v$ 
derived from the centroid shift of the emission lines, and the velocity dispersion $w$ derived from their 
broadening. In this work we applied our pipeline to a set of three hydrodynamical simulations 
\citep{Gaspari:2013_coma,Gaspari:2014_coma2} that model the injection of turbulence in a Coma-like 
galaxy cluster, assuming different Mach numbers \mnall. Using the \sixte\ software, we obtained 
three simulated event lists that represent X-IFU observations in ideal conditions (i.e.,~with enough 
photons not to be limited by statistics) and then applied a realistic observational pipeline to  
measure and map the quantities mentioned above. This was done by fitting about 600 spectra for each X-IFU FOV, allowing a 
comparison with the equivalent input projected quantities extracted directly from the 
hydrodynamical simulations. Finally, we computed the 2D structure functions of the three main 
quantities that can be used to describe the properties of turbulence, namely emission measure 
fluctuations $v$ and $w$, and verified our ability to recover their input slope and normalization.

Our main results can be summarized as follows:
\begin{enumerate}[\it i)]
\item The centroid shift and line broadening measured by fitting the X-IFU spectra  correspond well
to the emission-weighted velocity and velocity dispersion (see Eqs.~\ref{e:vew} and \ref{e:wew}, 
respectively) computed in the same region. The temperature is appropriately 
described by the spectroscopic-like temperature \citep[Eq.~\ref{e:tsl},][]{mazzotta04};
\item No significant bias is found in the measurement of the five physical quantities investigated in 
our analysis. The expected systematic uncertainties (listed in Table~\ref{t:res}) due to 
the modeling assumed in the fitting procedure are smaller than 5\% for all quantities, except 
for the broadening where it reaches about 15\%;
\item The physical quantities measured with our fitting procedure and their differences with 
the input projected values prove to be independent from each other at high statistical 
significance. This corroborates the accuracy of the current spectral fitting approach. Improvements 
on the fitting errors and to the residual systematic can be considered, for example using~optimal binning or 
other fitting methods;
\item The overall shape of the 2D structure functions of emission measure fluctuations, velocity, 
and velocity dispersion is  recovered well. However, we observe an excess of 15--30\% and of 
40--80\% in measurement of the normalization of SF$_v$ and SF$_w$, respectively: this is 
due to the limitations of the simplistic assumption on the error shapes;
\item We verified that these discrepancies do not affect the ability to measure the slope of the 
2D structure functions of all quantities and thus of their 2D power spectrum. In all cases the 
input slope is recovered at high precision, with differences smaller than 0.1.
\end{enumerate}

Our work highlights the unprecedented X-IFU capability of probing the thermodynamics and kinematics 
properties of the ICM, and of deriving the properties of its turbulent motions. In the near future 
we plan to extend our simulations to a larger FOV to better capture the injection scale of 
turbulence, and to run our simulator over a set of hydrodynamical simulations in a cosmological 
environment to estimate the cosmic variance associated with the turbulence power spectrum in a 
large-scale structure scenario. In addition, in the framework of \emph{Athena} science goals, it 
will be interesting to test the robustness of our fitting procedures using models that do not assume 
solar ratios for the metallicities, i.e.,~increasing the number of free parameters to measure 
 the abundance of the various elements independently.

\begin{acknowledgements}
This work has been completed despite the shameful situation of the Italian research system, worsened 
by an entire decade of severe cuts to the fundings of public universities and research institutes 
\cite[see, e.g.,][]{abbott06,abbott16,abbott18}. This has caused a whole generation of valuable 
researchers, in all fields, to struggle in poor working conditions with little hope of  achieving 
decent employment contracts and permanent positions, resulting in obvious difficulties in the 
planning of future research activities, and in open violation of the European Charter for 
Researchers\footnote{\href{https://euraxess.ec.europa.eu/jobs/charter}{https://euraxess.ec.europa.eu/jobs/charter}}. \\

We acknowledge support by the ASI (Italian Space Agency) through  Contract no. 2015-046-R.0. 
M.R. also acknowledges the financial contribution from   ASI agreement no. I/023/12/0 \emph{``Attivit\`a 
relative alla fase B2/C per la missione Euclid''}.
M.G. is supported by NASA through Einstein Postdoctoral Fellowship Award Number PF5-160137 issued by the 
Chandra X-ray Observatory Center, which is operated by the SAO for and on behalf of NASA under contract 
NAS8-03060. Support for this work was also provided by Chandra GO7-18121X. S.E. acknowledges financial 
contribution from  contracts NARO15 ASI-INAF I/037/12/0, ASI 2015-046-R.0, and ASI-INAF no. 2017-14-H.0. The
\texttt{FLASH} code was in part developed by the DOE NNSA-ASC OASCR Flash center at the University of Chicago.
HPC resources were in part provided by the NASA HEC Program (SMD-17-7251). E.R. acknowledges the ExaNeSt 
and Euro Exa projects, funded by the European Union’s Horizon 2020 research and innovation program 
under grant agreement No. 671553 and No. 754337 and financial contribution from   
ASI-INAF agreement no. 2017-14-H.0. \\

We thank the anonymous referee for providing useful comments. We are grateful to S.~Borgani, K.~Dolag, 
M.~Gitti, G.~Lanzuisi, and P.~Peille for insightful discussions. We also thank the \sixte\ development
team for their help and support.
\end{acknowledgements}

\bibliographystyle{aa} 
\bibliography{library.bib}


\begin{appendix}


\section{Statistical uncertainty with asymmetric errors}
\label{a:sstat}
The calculation of the expected variance of the difference between true and measured quantities in the case of 
asymmetric errors must take into account the properties of the split-normal (or double Gaussian) distribution 
\cite[see, e.g.,][]{wallis14}. Specifically, we consider that for a given measurement $\mu_{-\sigma_1}^{+\sigma_2}$ 
the posterior probability of the true value $X$ is described by the  distribution
\begin{equation}
f(X) = \left\{\begin{array}{l}
  A \, \exp{\left[-\frac{(X-\mu)^2}{2\sigma_1^2}\right]} \ \ \ {\rm for } \ X\leq \mu \vspace{0.2cm} \\
  A \, \exp{\left[-\frac{(X-\mu)^2}{2\sigma_2^2}\right]} \ \ \ {\rm for } \ X\geq \mu \, 
\end{array},\right.
\end{equation}
where $A=\left(\sqrt{2\pi}(\sigma_1+\sigma_2)/2\right)^{-1}$. In this case $\mu$ represents the mode of the distribution, 
i.e.,~the nominal value of the measurement. It is possible to show that the mean and the variance of the distributions are 
\begin{equation}
E(X) = \mu + \sqrt{\frac{2}{\pi}}(\sigma_1-\sigma_2)
\end{equation}
and
\begin{equation}
V(X) = \left(1-\frac{2}{\pi} \right)(\sigma_1-\sigma_2)^2 + \sigma_1 \sigma_2 \, ,
\end{equation}
respectively. Since we are interested in the expected dispersion with respect to $\mu$, which we consider as our reference value 
for the measurements, we can derive it from the last two equations,
\begin{equation}
\sigma_{\mu}^2(X)= V(X) + [E(X)-\mu]^2 = \left(\sigma_2-\sigma_1\right)^2 + \sigma_1 \sigma_2 \, , 
\end{equation}
which we adopt as a definition of $\sigma_{\rm stat}^2$ in Eq.~\ref{e:sstat}.


\section{Additional plots}
\label{a:scplots}
\begin{figure*}
\includegraphics[width=1.\textwidth]{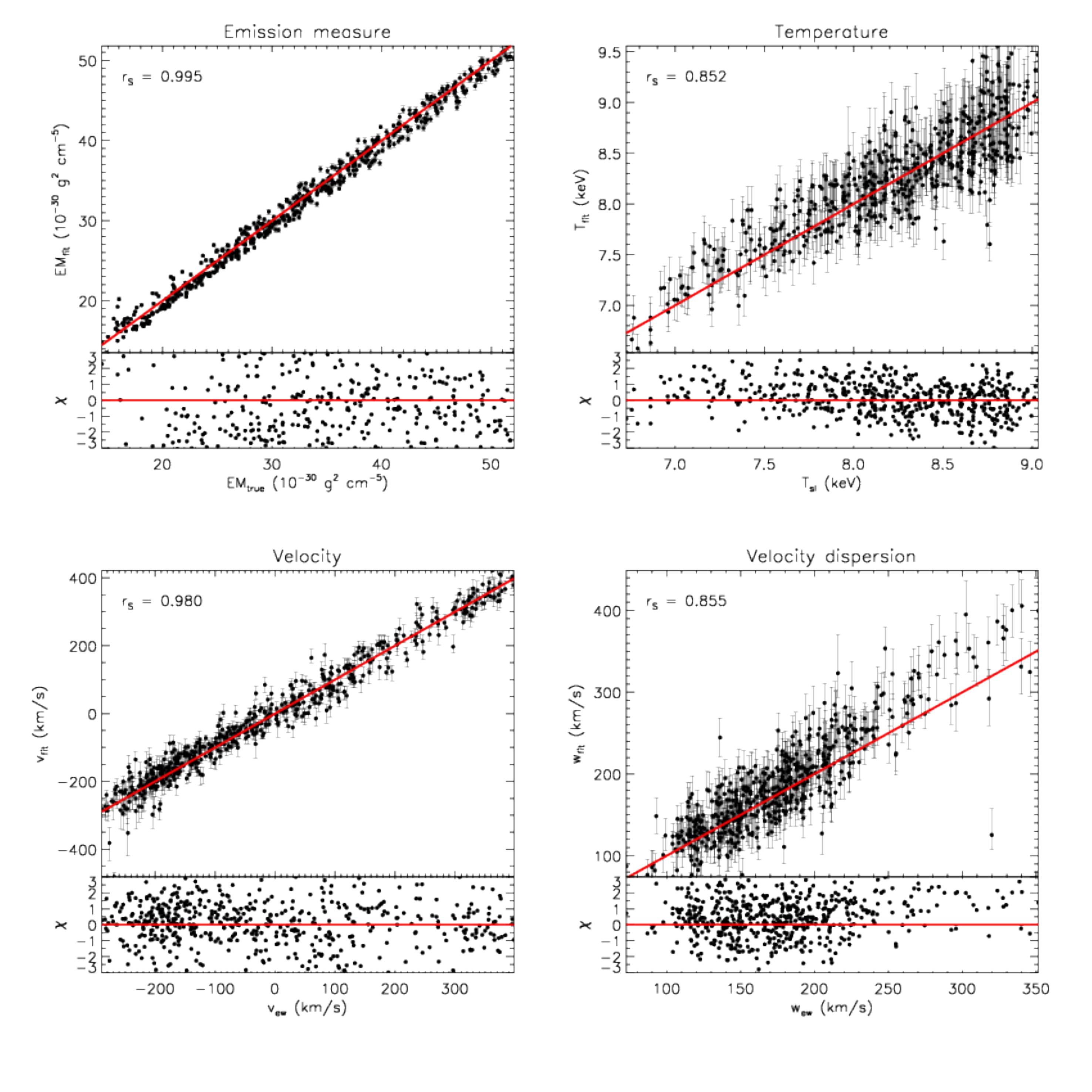}
\caption{
  Same as Fig.~\ref{f:scp}, but for the \mn\,=\,0.25 simulation.
  }
\label{f:scp025}
\end{figure*}

\begin{figure*}
\includegraphics[width=1.\textwidth]{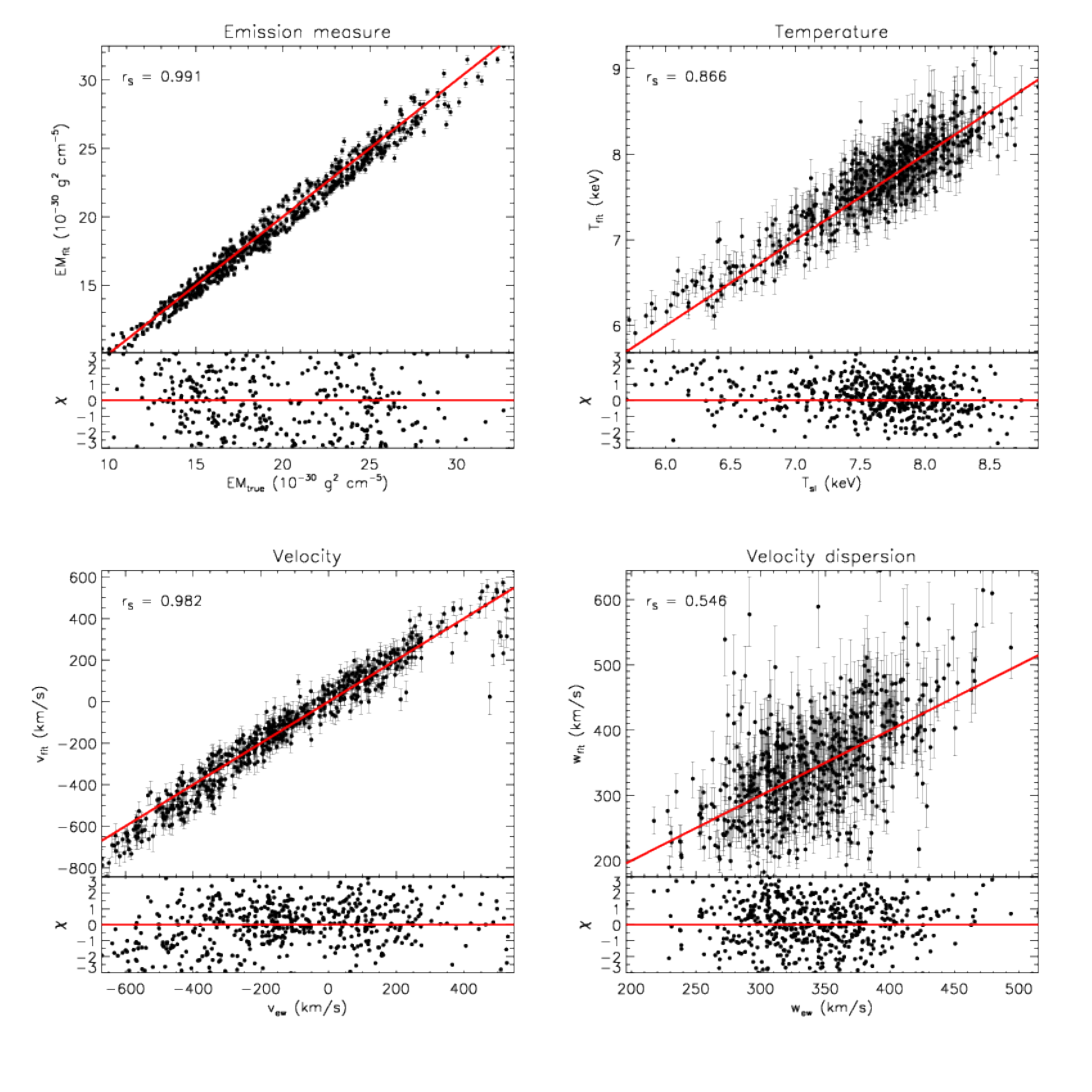}
\caption{
  Same as Fig.~\ref{f:scp}, but for the \mn\,=\,0.5 simulation.
  }
\label{f:scp05}
\end{figure*}

In this Appendix, as a further 
reference, we show the  plot in Fig.~\ref{f:scp} for the other two simulations. The plot for the \mn\,=\,0.25 and \mn\,=\,0.5 simulations are shown in Fig.~\ref{f:scp025} and Fig.~\ref{f:scp05},
respectively.

\label{lastpage}
\end{appendix}

\end{document}